\documentclass[11pt]{article}

\usepackage[tbtags]{amsmath}
\usepackage{amsthm}
\usepackage{amssymb}
\usepackage{amsfonts}
  \theoremstyle{plain}
  \newtheorem{theorem}{Theorem}
  \newtheorem{lemma}[theorem]{Lemma}
  \newtheorem{corollary}[theorem]{Corollary}
  \newtheorem{proposition}[theorem]{Proposition}
  
  \theoremstyle{definition}
  \newtheorem{definition}[theorem]{Definition}

  \theoremstyle{remark}
  \newtheorem*{remark}{Remark}
  \theoremstyle{plain}
  \newtheorem*{theorem*}{Theorem}
  \newtheorem*{lemma*}{Lemma}
  \newtheorem*{corollary*}{Corollary}
  \newtheorem*{proposition*}{Proposition}
  \newtheorem*{claim*}{Claim}

\newenvironment{step}
  {
    \begin{enumerate}

  }
  {\end{enumerate}}

\newenvironment{algorithm*}[1]
  {
    \begin{center}
      \hrulefill\\
      \textbf{#1}
  }
  {
    \vspace{-1\baselineskip}
    \hrulefill
    \end{center}
  }

\newenvironment{protocol*}[1]
  {
    \begin{center}
      \hrulefill\\
      \textbf{#1}
  }
  {
    \vspace{-1\baselineskip}
    \hrulefill
    \end{center}
  }

\newcommand{\bbC}{\mathbb{C}}

\newcommand{\bbN}{\mathbb{N}}

\newcommand{\bbZ}{\mathbb{Z}}

\newcommand{\bfD}{\mathbf{D}}

\newcommand{\bmM}{\boldsymbol{M}}

\newcommand{\calB}{\mathcal{B}}

\newcommand{\calE}{\mathcal{E}}

\newcommand{\calH}{\mathcal{H}}

\newcommand{\calK}{\mathcal{K}}

\newcommand{\calM}{\mathcal{M}}

\newcommand{\calR}{\mathcal{R}}
\newcommand{\calS}{\mathcal{S}}

\newcommand{\calV}{\mathcal{V}}

\newcommand{\sfB}{\mathsf{B}}

\newcommand{\sfH}{\mathsf{H}}

\newcommand{\sfR}{\mathsf{R}}
\newcommand{\sfS}{\mathsf{S}}

\newcommand{\sfV}{\mathsf{V}}

\newcommand{\classfont}{\mathrm}

\newcommand{\co}{\classfont{co}\textrm{-}}
\newcommand{\slashpoly}{\classfont{/poly}}
\newcommand{\slashqpoly}{\classfont{/qpoly}}
\newcommand{\NP}{\classfont{NP}}

\newcommand{\BPP}{\classfont{BPP}}
\newcommand{\PP}{\classfont{PP}}
\newcommand{\PSPACE}{\classfont{PSPACE}}
\newcommand{\EXP}{\classfont{EXP}}
\newcommand{\EXPTIME}{\classfont{EXPTIME}}

\newcommand{\PSPACEpoly}{\PSPACE\slashpoly}
\newcommand{\MA}{\classfont{MA}}
\newcommand{\QMA}{\classfont{QMA}}
\newcommand{\RQMA}{\classfont{RQMA}}
\newcommand{\EQMA}{\classfont{EQMA}}
\newcommand{\NQMA}{\classfont{NQMA}}
\newcommand{\QCMA}{\classfont{QCMA}}

\newcommand{\QMAqpoly}{\QMA\slashqpoly}
\newcommand{\NQP}{\classfont{NQP}}
\newcommand{\UP}{\classfont{UP}}
\newcommand{\coUP}{{\co\UP}}
\newcommand{\CequalP}{\classfont{C}_{=}\classfont{P}}
\newcommand{\coCequalP}{{\co\CequalP}}
\newcommand{\PH}{\classfont{PH}}

\newcommand{\tr}{\mathrm{tr}}

\newcommand{\bigprob}[1]{\Pr \bigl[ #1 \bigr]}

\newcommand{\expect}[1]{{\mathbf{E}[ #1 ]}}

\newcommand{\defeq}{\stackrel{\mathrm{def}}{=}}

\newcommand{\bra}[1]{\langle #1 \vert}

\newcommand{\ket}[1]{\vert #1 \rangle}

\newcommand{\ketbra}[1]{\vert #1 \rangle \langle #1 \vert}

\newcommand{\braket}[2]{\langle #1 \vert #2 \rangle}

\newcommand{\conjugate}[1]{#1^{\dagger}}

\newcommand{\normL}[1]{\left\Vert #1 \right\Vert}

\newcommand{\abs}[1]{\vert #1 \vert}

\newcommand{\function}[3]{{#1 \colon #2 \rightarrow #3}}

\newcommand{\bigset}[2]{{\bigl\{ #1 \colon #2 \bigr\}}}
\newcommand{\Bigset}[2]{{\Bigl\{ #1 \colon #2 \Bigr\}}}
\newcommand{\biggset}[2]{{\biggl\{ #1 \colon #2 \biggr\}}}
\newcommand{\Biggset}[2]{{\Biggl\{ #1 \colon #2 \Biggr\}}}

\newcommand{\Complex}{\bbC}

\newcommand{\Natural}{\bbN}
\newcommand{\Integers}{\bbZ}
\newcommand{\Nonnegative}{\Integers^{+}}

\newcommand{\Binary}{{\{ 0, 1 \}}}
\newcommand{\Density}{\bfD}

\newcommand{\acc}{\mathrm{acc}}

\newcommand{\yes}{\mathrm{yes}}
\newcommand{\no}{\mathrm{no}}
\newcommand{\Ayes}{A_{\yes}}
\newcommand{\Ano}{A_{\no}}

\begin{document}

\sloppy

% ---------------------------------------------------------------------------
%   Title page
% ---------------------------------------------------------------------------

\title{\Large
  \textbf{
    Quantum Merlin-Arthur Proof Systems:\\
    Are Multiple Merlins More Helpful to Arthur?\footnotemark[1]
  }\\
}

\author{
  Hirotada Kobayashi${}^{\text{a b }}$\footnotemark[2]\\
  \texttt{hirotada@nii.ac.jp}
  \and
  Keiji Matsumoto${}^{\text{a b }}$\footnotemark[3]\\
  \texttt{keiji@nii.ac.jp}
  \and
  Tomoyuki Yamakami${}^{\text{c }}$\footnotemark[4]\\
  \texttt{yamakami@u-aizu.ac.jp}
}

\renewcommand{\thefootnote}{\fnsymbol{footnote}}

\footnotetext[1]{
  A preliminary version of this paper has appeared in
  \emph{Algorithms and Computation, 14th International Symposium, ISAAC 2003},
  volume 2906 of \emph{Lecture Notes in Computer Science},
  pages 189--198, 2003~\cite{KobMatYam03ISAAC}.
}

\footnotetext[2]{
  Work partly done while at
  Department of Information Science,
  Graduate School of Science,
  The University of Tokyo
  and Quantum Computation and Information Project,
  Exploratory Research for Advanced Technology,
  Japan Science and Technology Agency.
  Partially supported by
  the Grant-in-Aid for Scientific Research~(B)~No.~18300002
  of the Ministry of Education, Culture, Sports, Science and Technology
  of Japan.
}  
\footnotetext[3]{
  Work partly done while at
  Quantum Computation and Information Project,
  Exploratory Research for Advanced Technology,
  Japan Science and Technology Agency.
  Partially supported by
  the Grant-in-Aid for Scientific Research~(B)~No.~18300002
  of the Ministry of Education, Culture, Sports, Science and Technology
  of Japan.
}  
\footnotetext[4]{
  Work partly done while at
  School of Information Technology and Engineering,
  University of Ottawa.
}

\date{}

\maketitle
\thispagestyle{empty}
\pagestyle{plain}
\setcounter{page}{0}

\vspace{-5mm}

\renewcommand{\thefootnote}{\alph{footnote}}

\begin{center}
{\large
  \footnotemark[1]%
  Principles of Informatics Research Division\\
  National Institute of Informatics\\
%%   Tokyo, Japan\\
  2-1-2 Hitotsubashi, Chiyoda-ku, Tokyo 101-8430, Japan\\
  [2.5mm]
  \footnotemark[2]%
  Quantum Computation and Information Project\\
  Solution Oriented Research for Science and Technology\\
  Japan Science and Technology Agency\\
%%   Tokyo, Japan\\
  5-28-3 Hongo, Bunkyo-ku, Tokyo 113-0033, Japan\\
  [2.5mm]
  \footnotemark[3]%
  School of Computer Science and Engineering\\
  The University of Aizu\\
%%   Aizu-Wakamatsu, Fukushima, Japan\\
  90 Kami-Iawase, Tsuruga, Ikki-machi, Aizu-Wakamatsu, Fukushima 965-8580, Japan
}\\
[5mm]
{\large 12 May 2008}\\
[8mm]
\end{center}

\renewcommand{\thefootnote}{\arabic{footnote}}

% ---------------------------------------------------------------------------
%   Abstract
% ---------------------------------------------------------------------------

\begin{abstract}
This paper introduces quantum ``multiple-Merlin''-Arthur proof systems
in which Arthur receives multiple quantum proofs
that are unentangled with each other.
Although classical multi-proof systems
are obviously equivalent to classical single-proof systems
(i.e., usual Merlin-Arthur proof systems),
it is unclear whether or not quantum multi-proof systems
collapse to quantum single-proof systems
(i.e., usual quantum Merlin-Arthur proof systems).
This paper presents a necessary and sufficient condition
under which the number of quantum proofs is reducible to two.
It is also proved that,
in the case of perfect soundness,
using multiple quantum proofs does not increase
the power of quantum Merlin-Arthur proof systems.
\end{abstract}

\clearpage

% ---------------------------------------------------------------------------
%   Introduction
% ---------------------------------------------------------------------------

\section{Introduction}
\label{Section: introduction}

% ---------------------------------------------------------------------------
%   Background
% ---------------------------------------------------------------------------

\subsection{Background}
\label{Subsection: background}

Merlin-Arthur proof systems,
or Merlin-Arthur games as originally called,
were introduced by Babai~\cite{Bab85STOC}.
In a Merlin-Arthur proof system,
powerful Merlin, a prover, presents a proof
and Arthur, a verifier, probabilistically verifies its correctness
with high success probability.
The class of problems having Merlin-Arthur proof systems is denoted by $\MA$,
and has played important roles
in computational complexity theory~\cite{Bab85STOC, BabMor88JCSS, BabForLun91CC, BabForNisWig93CC, Bab92JDM, Ver92SCT, RusSun98CC, ArvKob01TCS, GolZuc97ECCC, BuhMelRegSivStr00SIComp, BuhMel99JCSS, ImpKabWig02JCSS, San07STOC}.

A quantum analogue of $\MA$ was first discussed
by Knill~\cite{Kni96TR}
and has been studied intensively~\cite{Kit99AQIP, Wat00FOCS, KemReg03QIC, JanWocBet05IJQI, WocJanBet03QIC, AhaReg03FOCS, MarWat05CC, KemKitReg06SIComp, Aar06CCC, Liu06RANDOM-APPROX, LiuChrVer07PRL, AarKup07ToC, AhaGotIraKem07FOCS, Liu07PhD}.
%% both before the completion of this work~\cite{Kit99AQIP, Wat00FOCS, KemReg03QIC, JanWocBet05IJQI, WocJanBet03QIC, AhaReg03FOCS}
%% and after that~\cite{MarWat05CC, KemKitReg06SIComp, Aar06CCC, Liu06RANDOM-APPROX, LiuChrVer07PRL, AarKup07ToC, AhaGotIraKem07FOCS, Liu07PhD}.
In the most commonly-used version of
quantum Merlin-Arthur proof systems,
a proof presented by Merlin is a pure quantum state
called a \emph{quantum proof}
and Arthur's verification process is a polynomial-time quantum computation.
However, all the previous studies only consider the model
in which Arthur receives a single quantum proof,
and no discussions are done so far on the model
in which Arthur receives \emph{multiple} quantum proofs unentangled
with each other.

Classically, multiple proofs can be concatenated into a long single proof,
and thus, there is no advantage to use multiple proofs.
Quantumly, however, using multiple quantum proofs
may not be computationally equivalent to using a single quantum proof,
because knowing that a given proof is a tensor product
of some quantum states might be advantageous to Arthur.
For example, in the case of two quantum proofs versus one,
consider the following most straightforward Arthur's simulation
of two quantum proofs by a single quantum proof:
given a single quantum proof that is expected to be
a tensor product of two pure quantum states,
Arthur first runs some preprocessing to rule out any quantum proof
far from states of a tensor product of two pure quantum states,
and then performs the verification procedure of the original two-proof system.
It turns out that this most straightforward method does not work well,
since there is no physical method
that determines whether a given unknown state is in a tensor product form
or even maximally entangled, as will be shown in Section~\ref{Section: discussions}.
Another fact is that the unpublished proof by Kitaev~and~Watrous
for the upper bound $\PP$
of the class $\QMA$ of problems having
single-proof quantum Merlin-Arthur proof systems
no longer works well for the multi-proof cases
with the most straightforward modification.
The simplified proof by Marriott~and~Watrous~\cite{MarWat05CC} for the same statement
and even the proof of
${\QMA \subseteq \PSPACE}$~\cite{Kit99AQIP, KitSheVya02Book}
are also the cases.
Furthermore, the existing proofs for the property
that parallel repetition of a single-proof system
reduces the error probability to be arbitrarily small~\cite{KitWat00STOC, Wat00FOCS, KitSheVya02Book, MarWat05CC}
cannot be applied to the multi-proof cases.
Of course, these arguments do not imply that
using multiple quantum proofs is more powerful
than using only a single quantum proof
from the complexity theoretical viewpoint.
The authors believe, however, that these at least justify that
it is meaningful to consider the multi-proof model of
quantum Merlin-Arthur proof systems.
It is interesting to note that here the \emph{nonexistence} of entanglement
among proofs may have the possibility of enhancing the verification power,
unlike the usual situations of quantum information processing where
we make use of the \emph{existence} of entanglement.
Moreover, the multi-proof model has importance
even in quantum information theory,
because the model is inherently related to entanglement theory.
Indeed, after the completion of this work,
Aaronson,~Beigi,~Drucker,~Fefferman,~and~Shor~\cite{AarBeiDruFefSho08CCC}
succeeded in proving a strong connection between our model
and the famous ``Additivity Conjecture'' in entanglement theory,
which is one of the most important conjectures in quantum information theory.

% ---------------------------------------------------------------------------
%   Contribution of This Paper
% ---------------------------------------------------------------------------

\subsection{Contribution of This Paper}
\label{Subsection: contribution}

Motivated by the observations listed in the previous subsection,
this paper extends the usual single-proof model of
quantum Merlin-Arthur proof systems
to the multi-proof model by allowing Arthur to use multiple quantum proofs,
which are given in a tensor product form of multiple quantum states.
One may think of this model as
a special case of
quantum multi-prover interactive proof systems~\cite{KobMat03JCSS}
in which a verifier cannot ask questions to provers,
and provers do not share entanglement a priori.
Formally, we say that a problem ${A = \{\Ayes, \Ano\}}$ has
a \emph{${(k, c, s)}$-quantum Merlin-Arthur proof system}
if there exists a polynomial-time quantum verifier $V$ such that,
for every input $x$,
(i) if ${x \in \Ayes}$, there exists a set of ${k(\abs{x})}$ quantum proofs
that makes $V$ accept $x$ with probability at least ${c(\abs{x})}$,
and (ii) if ${x \in \Ano}$, for any set of ${k(\abs{x})}$ quantum proofs given,
$V$ accepts $x$ with probability at most ${s(\abs{x})}$.
The resulting complexity class is denoted by ${\QMA(k, c, s)}$.
We often abbreviate ${\QMA \bigl(k, \frac{2}{3}, \frac{1}{3}\bigr)}$
as ${\QMA(k)}$ throughout this paper.\footnotemark

\footnotetext{
  Here we choose completeness and soundness accepting probabilities
  $\frac{2}{3}$ and $\frac{1}{3}$ to define the class ${\QMA(k)}$,
  but there may be other reasonable choices.
  For instance, ${\QMA(k)}$ could be defined as
  the union of ${\QMA(k, 1 - \varepsilon, \varepsilon)}$
  for all negligible functions $\varepsilon$.
%%   the intersection of ${\QMA(k, c, s)}$
%%   for all two-sided bounded error probabilities ${(c,s)}$.
  It is possible that other reasonable definitions of ${\QMA(k)}$
  form different classes from the one defined in this paper,
  since it is not known how to amplify the success probability of ${\QMA(k)}$.
  The authors believe, however, that
  the choice of $\frac{2}{3}$ and $\frac{1}{3}$
  would best highlight the essence of the results in this paper.
}

Besides our central question
whether or not
quantum multi-proof Merlin-Arthur proof systems collapse to
quantum single-proof systems,
it is also unclear
if there are $k_1$ and $k_2$ with ${k_1 \neq k_2}$
such that ${\QMA(k_1) = \QMA(k_2)}$.
Towards settling these questions,
this paper presents
a necessary and sufficient condition under which
the number of quantum proofs is reducible to two.
Our condition is related to
the possibility of amplifying success probability
of quantum two-proof Merlin-Arthur proof systems
without increasing the number of quantum proofs.
More formally,
it is proved that
${\QMA(k, c, s) = \QMA \bigl(2, \frac{2}{3}, \frac{1}{3} \bigr)}$
for any polynomially-bounded function ${k \geq 2}$
and any two-sided bounded error probability ${(c, s)}$
if and only if
${\QMA(2, c, s) = \QMA \bigl(2, \frac{2}{3}, \frac{1}{3} \bigr)}$
for any two-sided bounded error probability ${(c, s)}$.\footnotemark
\footnotetext{
  This improves the result
  proved in our preliminary conference version~\cite{KobMatYam03ISAAC},
  where we required the amplifiablity of the success probability
  not only for two-proof systems
  but also for $k$-proof systems, for every $k$.
  Also, the statement was originally proved only for every \emph{constant} $k$,
  whereas the improved statement in the current version holds even for
  every polynomially-bounded function $k$.
  The same improvements were done independently by
  Aaronson,~Beigi,~Drucker,~Fefferman,~and~Shor~\cite{AarBeiDruFefSho08CCC}.
}
Alternatively, it is proved that
quantum multi-proof Merlin-Arthur proof systems
are equivalent to usual single-proof ones
if and only if
quantum \emph{two-proof} Merlin-Arthur proof systems
are equivalent to usual single-proof ones.
That is,
${\QMA(k, c, s) = \QMA}$
for any polynomially-bounded function ${k \geq 2}$
and any two-sided bounded error probability ${(c, s)}$
if and only if
${\QMA(2, c, s) = \QMA}$
for any two-sided bounded error probability ${(c, s)}$.
The key ingredient to show these properties
is the claim that,
for any quantum multi-proof Merlin-Arthur proof system
with some appropriate condition on completeness and soundness,
we can reduce the number of proofs by (almost) two-thirds
(where the gap between completeness and soundness becomes worse,
but is still bounded by an inverse-polynomial).
This is done by using the controlled-swap test,
which often plays a key role in quantum computation
(e.g., in Refs.~\cite{KitWat00STOC, BuhCleWatWol01PRL}).

It is also proved for the case of perfect soundness
that,
for any polynomially-bounded function ${k \geq 2}$
and any completeness $c$,
${\QMA(k, c, 0) = \QMA(1, c, 0)}$.
With further analyses,
the class $\NQP$,
which derives from another concept of ``quantum nondeterminism''
introduced by Adleman,~DeMarrais,~and~Huang~\cite{AdlDeMHua97SIComp}
and discussed by a number of studies~\cite{ForRog99JCSS, FenGreHomPru99PRSLA, YamYao99IPL, Wol03SIComp},
is characterized by the union of ${\QMA(1, c, 0)}$
for all error probability functions $c$.
This bridges between two existing concepts of ``quantum nondeterminism''.

% ---------------------------------------------------------------------------
%   Recent Progresses
% ---------------------------------------------------------------------------

\subsection{Recent Progresses}
\label{Subsection: recent progresses}

After the completion of this work,
a number of studies showed very intriguing properties on our model.

Liu,~Christandl,~and~Verstraete~\cite{LiuChrVer07PRL}
showed that the \textsc{Pure State $N$-Representability} problem,
which naturally arises in quantum chemistry, can be verified
by a quantum two-proof Merlin-Arthur proof system
with two-sided bounded error.
Interestingly, the problem is not known to be in usual $\QMA$.

Blier~and~Tapp~\cite{BliTap07arXiv}
proved that the $\NP$-complete problem \textsc{Graph 3-Coloring}
has a quantum two-proof Merlin-Arthur proof system
with one-sided bounded error of perfect completeness,
where both of the two unentangled quantum proofs
consist of only logarithmically many qubits.
The soundness is bounded away from one only by an inverse-polynomial
in their proof system.

Aaronson,~Beigi,~Drucker,~Fefferman,~and~Shor~\cite{AarBeiDruFefSho08CCC}
proved that the $\NP$-complete problem \textsc{3-SAT}
has a quantum multi-proof Merlin-Arthur proof system
of perfect completeness with constant soundness error,
where the number of proofs is almost square root of the instance size,
and each quantum proof consists of only logarithmically many qubits.
They further showed that
the ``Additivity Conjecture'' would imply that
any quantum two-proof Merlin-Arthur proof system
can be made to have arbitrarily small two-sided bounded error,
and thus, ${\QMA(k) = \QMA(2)}$ for any polynomially bounded function
${k \geq 2}$.

% ---------------------------------------------------------------------------
%   Organization of This Paper
% ---------------------------------------------------------------------------

\subsection{Organization of This Paper}
\label{Subsection: organization}

The remainder of this paper is organized as follows.
In Section~\ref{Section: preliminaries}
we give a brief review for several basic notions
of quantum computation and information theory used in this paper.
In Section~\ref{Section: definitions}
we formally define
the multi-proof model of quantum Merlin-Arthur proof systems. 
In Section~\ref{Section: condition under which QMA(k) = QMA(2)}
we show a condition under which ${\QMA(k) = \QMA(2)}$.
In Section~\ref{Section: cases with perfect soundness}
we focus on the systems of perfect soundness.
In Section~\ref{Section: discussions}
we show that there is no physical method
that determines whether a given unknown state is in a tensor product form
or maximally entangled.
Finally, we conclude with Section~\ref{Section: conclusions}
which summarizes this paper.
The conference version of this paper~\cite{KobMatYam03ISAAC}
also included the result that
there exists an oracle relative to which ${\QMA(k)}$ does not contain $\coUP$.
The present version omits this result,
since it turned out that the statement is easily proved by using
the result by Raz~and~Shpilka~\cite{RazShp04CCC}.

% ---------------------------------------------------------------------------
%   Preliminaries
% ---------------------------------------------------------------------------

\section{Preliminaries}
\label{Section: preliminaries}

We start with reviewing several fundamental notions used in this paper.
Throughout this paper
we assume that
all input strings are over the alphabet ${\Sigma = \{ 0, 1 \}}$,
and $\Natural$ and $\Nonnegative$
denote the sets of positive and nonnegative integers, respectively.
For any Hilbert space $\calH$,
let $I_{\calH}$ denote the identity operator over $\calH$.
In this paper, all Hilbert spaces are of dimension power of two.

% ---------------------------------------------------------------------------
%   Quantum Fundamentals
% ---------------------------------------------------------------------------

\subsection{Quantum Fundamentals}
\label{Subsection: quantum fundamentals}

First we briefly review basic notations and definitions
in quantum computation and quantum information theory.
Detailed descriptions are found in Refs.~\cite{NieChu00Book, KitSheVya02Book},
for instance.

A \emph{pure quantum state}, or a \emph{pure state} in short,
is a unit vector $\ket{\psi}$ in some Hilbert space $\calH$.
For any Hilbert space $\calH$,
let $\ket{0_{\calH}}$ denote the pure quantum state in $\calH$
of which all the qubits are in state $\ket{0}$.
A \emph{mixed quantum state}, or a \emph{mixed state} in short,
is a classical probability distribution
${(p_{i}, \ket{\psi_i})}$, ${0 \leq p_i \leq 1}$, ${\sum_i p_i = 1}$
over pure states ${\ket{\psi_i} \in \calH}$.
This can be interpreted as being in the pure state
$\ket{\psi_i}$ with probability $p_i$.
A mixed state ${(p_i, \ket{\psi_i})}$
is often described in the form of a \emph{density operator}
${\rho = \sum_i p_i \ketbra{\psi_i}}$.
Any density operator is positive semidefinite and has trace one.
It should be noted that different probabilistic mixtures of pure states
can yield mixed states with the identical density operator.
It is also noted that there is no physical method (i.e., no measurement)
to distinguish mixed states with the identical density operator.
Therefore, density operators give complete descriptions of quantum states,
and we may use the term ``density operator''
to indicate the corresponding mixed state.
For any Hilbert space $\calH$,
let ${\Density(\calH)}$ denote the set of density operators over $\calH$.

One of the important operations to density operators is
the \emph{trace-out} operation.
Given Hilbert spaces $\calH$ and $\calK$
and a quantum state with its density operator $\rho$ in ${\Density(\calH \otimes \calK)}$,
the quantum state after \emph{tracing out} $\calK$
has its density operator in ${\Density(\calH)}$ defined by
${
  \tr_{\calK} \rho
  =
  \sum_{i=1}^d (I_{\calH} \otimes \bra{e_i}) \rho (I_{\calH} \otimes \ket{e_i})
}$
for any orthonormal basis ${\{ \ket{e_i} \}}$ of $\calK$,
where $d$ is the dimension of $\calK$.
To perform this operation on some part of a quantum system
gives a partial view of the quantum system with respect to the remaining part.

A \emph{positive operator-valued measure~(POVM)} on a Hilbert space $\calH$
is defined to be a set ${\bmM = \{M_1, \ldots, M_k\}}$
of nonnegative Hermitian operators over $\calH$
such that ${\sum_{i=1}^k M_i = I_{\calH}}$.
For any POVM $\bmM$ on $\calH$,
there is a quantum mechanical measurement
that results in $i$ with probability exactly $\tr (M_i \rho)$
for any $\rho$ in ${\Density(\calH)}$.
See Refs.~\cite{Hol82Book, Oza84JMathPhy}
for more rigorous descriptions on quantum measurements.

The \emph{fidelity} ${F(\rho, \sigma)}$
between two density operators $\rho$ and $\sigma$ in ${\Density(\calH)}$
is defined by
${F(\rho, \sigma) = \tr \sqrt{\sqrt{\rho} \sigma \sqrt{\rho}}}$.
This paper uses the following two properties on fidelity.

\begin{lemma}[\cite{Joz94JModOpt}]
For any Hilbert spaces $\calH$ and $\calK$
and any density operators ${\rho_1, \sigma_1 \in \Density(\calH)}$
and ${\rho_2, \sigma_2 \in \Density(\calK)}$,
\[
F(\rho_{1} \otimes \rho_{2}, \sigma_{1} \otimes \sigma_{2})
=
F(\rho_{1}, \sigma_{1})F(\rho_{2}, \sigma_{2}).
\]
\label{Lemma: product of fidelity}
\end{lemma}

\begin{lemma}[\cite{SpeRud02PRA,NaySho03PRA}]
For any Hilbert space $\calH$
and any density operators ${\rho, \sigma, \xi \in \Density(\calH)}$,
%% ${F(\rho, \sigma)^2 + F(\sigma, \xi)^2 \leq 1 + F(\rho, \xi)}$.
\[
F(\rho, \sigma)^2 + F(\sigma, \xi)^2
\leq
1 + F(\rho, \xi).
\]
\label{Lemma: F(a,b)^2 + F(b,c)^2 < 1 + F(a,c)}
\end{lemma}

% ---------------------------------------------------------------------------
%   Quantum Circuits
% ---------------------------------------------------------------------------

\subsection{Quantum Circuits}
\label{Subsection: quantum circuits}

Next we review the model of quantum circuits.
We use the following notion of
polynomial-time uniformly generated families of quantum circuits.

A quantum circuit consists of a finite number of quantum gates
that are applied in sequence to a finite number of qubits.
A family ${\{ Q_x \}}$ of quantum circuits is
\emph{polynomial-time uniformly generated}
if there exists a deterministic procedure
that, on every input $x$, outputs a description of $Q_x$
and runs in time polynomial in $\abs{x}$.
It is assumed that the circuits in such a family are composed of gates
in some reasonable, universal, finite set of quantum gates.
Furthermore, it is assumed that the number of gates in any circuit
is not more than the length of the description of that circuit.
Therefore $Q_x$ must have size polynomial in $\abs{x}$.
For convenience,
we may identify a circuit $Q_x$ with the unitary operator it induces.

Since non-unitary and unitary quantum circuits
are equivalent in computational power~\cite{AhaKitNis98STOC},
it is sufficient to treat only unitary quantum circuits,
which justifies the above definition.
For avoiding unnecessary complication, however,
the descriptions of procedures may include non-unitary operations
in the subsequent sections.
Even in such cases, it is always possible to construct
unitary quantum circuits
that essentially achieve the same procedures described.

% ---------------------------------------------------------------------------
%   Definitions
% ---------------------------------------------------------------------------

\section{Definitions}
\label{Section: definitions}

Here we formally define quantum \emph{multi-proof} Merlin-Arthur proof systems.
Although all the statements in this paper
can be proved only in terms of languages
without using promise problems~\cite{EveSelYac84InfoCont},
in what follows we define models and prove statements
in terms of promise problems,
for generality
and for the compatibility with some subsequent studies
on our model~\cite{LiuChrVer07PRL, Liu07PhD}.
%% and for the compatibility with some other studies
%% on quantum Merlin-Arthur proof systems~\cite{Kit99AQIP, KemReg03QIC, JanWocBet05IJQI, WocJanBet03QIC, AhaReg03FOCS, KemKitReg06SIComp, Liu06RANDOM-APPROX, LiuChrVer07PRL, AhaGotIraKem07FOCS, Liu07PhD}.

A \emph{quantum proof of size $q$} is a pure quantum state of $q$ qubits.

A \emph{quantum verifier $V$ for quantum $k$-proof Merlin-Arthur proof systems}
is a polynomial-time computable mapping of the form
$\function{V}{\Sigma^{\ast}}{\Sigma^{\ast}}$.
For every input ${x \in \Sigma^{\ast}}$,
the string ${V(x)}$ is interpreted as
a description of a polynomial-size quantum circuit.
In other words, ${\{ V(x) \}}$ forms a polynomial-time uniformly generated
family of quantum circuits.
The qubits upon which each ${V(x)}$ acts are divided into ${k+1}$ sets:
one set, consisting of ${q_{\calV}(\abs{x})}$ qubits,
serves as work space of $V$,
and each of the rest $k$ sets
serves as ``witness space'' of $V$
that is used for storing a quantum proof of size ${q_{\calM}(\abs{x})}$,
for some polynomially bounded functions
$\function{q_{\calV}, q_{\calM}}{\Nonnegative}{\Natural}$.
One of the qubits in the work space of $V$ is designated as the output qubit.

A set of $k$ quantum proofs is \emph{compatible} with a quantum verifier $V$
if the size of every quantum proof coincides with
the size of witness space of $V$.

Suppose that $V$ receives $k$ quantum proofs ${\ket{\phi_1}, \ldots, \ket{\phi_k}}$.
The probability that $V$ accepts $x$ is defined
to be the probability that an observation of the output qubit
in the ${\{ \ket{0}, \ket{1} \}}$ basis yields $\ket{1}$,
after the circuit ${V(x)}$ is applied to the state
${
  \ket{0_{\calV}}
  \otimes
  \ket{\phi_1} \otimes \cdots \otimes \ket{\phi_k}
}$,
where $\calV$ is the Hilbert space
corresponding to the work space of $V$.
%% ${
%%   \ket{0^{q_{\calV}(\abs{x})}}
%%   \otimes
%%   \ket{\phi_1} \otimes \cdots \otimes \ket{\phi_k}
%% }$.

More generally, the number of quantum proofs
may not necessarily be a constant,
and may be a polynomially bounded function
$\function{k}{\Nonnegative}{\Natural}$ of the input length.

\begin{definition}
Given a polynomially bounded function $\function{k}{\Nonnegative}{\Natural}$
and functions $\function{c,s}{\Nonnegative}{[0,1]}$,
a problem ${A = \{ A_{\yes}, A_{\no} \}}$ is in ${\QMA(k, c, s)}$
if there exists a quantum verifier $V$
for $k$-proof quantum Merlin-Arthur proof systems
such that, for every $x$,
\begin{description}
\item[\textnormal{(Completeness)}]
if ${x \in \Ayes}$,
there exists a set of quantum proofs
${\ket{\phi_1}, \ldots, \ket{\phi_{k(\abs{x})}}}$
compatible with $V$
that makes
$V$ accept $x$ with probability at least ${c(\abs{x})}$,
\item[\textnormal{(Soundness)}]
if ${x \in \Ano}$,
for any set of quantum proofs
${\ket{\phi_1}, \ldots, \ket{\phi_{k(\abs{x})}}}$
compatible with $V$,
$V$ accepts $x$ with probability at most ${s(\abs{x})}$.
\end{description}
\label{Definition: QMA(k, c, s)}
\end{definition}

We say that a problem ${A = \{ A_{\yes}, A_{\no} \}}$
has a \emph{${(k, c, s)}$-quantum Merlin-Arthur proof system},
or a \emph{${\QMA(k, c, s)}$ proof system} in short,
if and only if $A$ is in ${\QMA(k, c, s)}$.
For simplicity,
we abbreviate ${\QMA \bigl(k, \frac{2}{3}, \frac{1}{3}\bigr)}$
as ${\QMA(k)}$ for every $k$.

Note that allowing quantum proofs of mixed states
does not increase the maximal accepting probability of proof systems,
which justifies the model defined above.
For readability, in what follows,
the arguments $x$ and $\abs{x}$ may be dropped in various functions,
if it is not confusing.

% ---------------------------------------------------------------------------
%   Condition under which QMA(k) = QMA(2)
% ---------------------------------------------------------------------------

\section{\boldmath{Condition under which ${\QMA(k) = \QMA(2)}$}}
\label{Section: condition under which QMA(k) = QMA(2)}

Classically, it is trivial to show
that classical multi-proof Merlin-Arthur proof systems
are essentially equivalent to single-proof ones.
However, it is unclear
whether quantum multi-proof Merlin-Arthur proof systems collapse to
quantum single-proof systems.
Moreover, it is also unclear
whether there are $k_1$ and $k_2$ of ${k_1 \neq k_2}$
such that ${\QMA(k_1) = \QMA(k_2)}$.
Towards settling these questions,
here we give a condition under which ${\QMA(k) = \QMA(2)}$
for every polynomially-bounded function ${k \geq 2}$.

Formally, we consider the following condition
on the possibility of amplifying the success probability
of quantum two-proof Merlin-Arthur proof systems
without increasing the number of quantum proofs:
\begin{itemize}
  \item[($\ast$)]
    For any two-sided bounded error probability ${(c, s)}$,
    ${\QMA(2, c, s)}$ coincides with
    ${\QMA \bigl(2, \frac{2}{3}, \frac{1}{3} \bigr)}$.
\end{itemize}
Our main result is the following theorem, which will be shown in this section.

\begin{theorem}
${\QMA(k, c, s) = \QMA \bigl(2, \frac{2}{3}, \frac{1}{3} \bigr)}$
for any polynomially-bounded function
$\function{k}{\Nonnegative}{\Natural}$ satisfying ${k \geq 2}$
and any two-sided bounded error probability ${(c, s)}$
if and only if the condition ($\ast$) is satisfied.
\label{Theorem: necessary and sufficient condition for QMA(k, c, s) = QMA(2, 2/3, 1/3)}
\end{theorem}

% ---------------------------------------------------------------------------
%   Achieving Exponentially Small Completeness Error
% ---------------------------------------------------------------------------

\subsection{Achieving Exponentially Small Completeness Error}

We first show a simple way
of achieving exponentially small completeness error
while keeping soundness error bounded away from one,
which works well for \emph{any} proof systems.
%% including quantum multi-proof Merlin-Arthur proof systems.
The same result was independently proved by
Aaronson,~Beigi,~Drucker,~Fefferman,~and~Shor~\cite[Lemma~6]{AarBeiDruFefSho08CCC}.

\begin{lemma}
Let $\function{c, s}{\Nonnegative}{[0,1]}$
be any functions that satisfy
${c - s \geq \frac{1}{q}}$
for some polynomially bounded function
$\function{q}{\Nonnegative}{\Natural}$,
and let $\Pi$ be any proof system
with completeness at least $c$ and soundness at most $s$.
Consider another proof system $\Pi'$
such that,
for every input of length $n$,
$\Pi'$ carries out ${N = 2 p(n) (q(n))^2}$ attempts of $\Pi$ in parallel
for a polynomially bounded function
$\function{p}{\Nonnegative}{\Natural}$,
and accepts
iff at least ${\frac{c(n)+s(n)}{2}}$-fraction of these $N$ attempts
results in acceptance in $\Pi$.
Then $\Pi'$
has completeness at least ${1 - 2^{-p}}$
and soundness at most
${\frac{2s}{c+s} \leq 1 - \frac{c-s}{2} \leq 1 - \frac{1}{2q}}$.
\label{Lemma: exp-small completeness error}
\end{lemma}

\begin{proof}
Let $X_i$ be the random variable
that takes $1$ iff the $i$th attempt of $\Pi$ in $\Pi'$
results in acceptance and otherwise takes $0$,
for each ${1 \leq i \leq N}$,
and let $Y$ be the random variable
defined by ${Y = \frac{\sum_{i=1}^N X_i}{N}}$.

Noticing that
the accepting probability in $\Pi'$
is given by
$\bigprob{Y \geq \frac{c(n)+s(n)}{2}}$
and that
${\expect{Y} = \frac{\sum_{i=1}^N \expect{X_i}}{N}}$,
the completeness bound of $\Pi'$
directly follows from the Hoeffding bound
while the soundness bound of $\Pi'$
directly follows from Markov's inequality.
\end{proof}

The following is an immediate corollary of 
Lemma~\ref{Lemma: exp-small completeness error}.

\begin{corollary}
For any polynomially bounded functions
$\function{k,p}{\Nonnegative}{\Natural}$
and any two-sided bounded error probability ${(c,s)}$,
\[
\QMA(k, c, s)
\subseteq
\QMA \Bigl( k, 1 - 2^{-p}, \frac{2s}{c+s} \Bigr)
\subseteq
\QMA \Bigl( k, 1 - 2^{-p}, 1 - \frac{c-s}{2} \Bigr).
\]
\label{Corollary: exp-small completeness error for QMA}
\end{corollary}

% ---------------------------------------------------------------------------
%   A Key Lemma
% ---------------------------------------------------------------------------

\subsection{Controlled-Swap Test with Mixed States}

Next we show a fundamental property of the controlled-swap test
when applied to a pair of mixed states.
This property is easy to prove.
To the best knowledge of the authors, however,
it has not appeared previously,
and the authors believe that this property will be useful in many cases.

The controlled-swap operator exchanges
the contents of two registers $\sfR_1$ and $\sfR_2$
if the control register $\sfB$ contains $1$,
and does nothing if $\sfB$ contains $0$.

Given a pair of mixed states $\rho$ and $\sigma$ of $n$ qubits
of the form ${\rho \otimes \sigma}$,
prepare quantum registers $\sfB$, $\sfR_1$, and $\sfR_2$.
The register $\sfB$ consists of only one qubit
that is initially set to state $\ket{0}$,
while the registers $\sfR_1$ and $\sfR_2$ consist of $n$ qubits
and $\rho$ and $\sigma$ are initially set in $\sfR_1$ and $\sfR_2$,
respectively.

The controlled-swap test is performed
by running the algorithm described in Figure~\ref{Figure: controlled-swap test}.

\begin{figure}
\begin{algorithm*}{Controlled-Swap Test}
\begin{step}
\item
  Apply the Hadamard transformation $H$ to $\sfB$.
\item
  Apply the controlled-swap operator to $\sfR_1$ and $\sfR_2$
  using $\sfB$ as a control qubit.
  That is, 
  swap the contents of $\sfR_1$ and $\sfR_2$
  if $\sfB$ contains $1$,
  and do nothing if $\sfB$ contains $0$. 
\item
  Apply the Hadamard transformation $H$ to $\sfB$.
  Accept if $\sfB$ contains $0$, and reject otherwise.
\end{step}
\end{algorithm*}
\caption{The controlled-swap test.}
\label{Figure: controlled-swap test}
\end{figure}

\begin{proposition}
The probability that the input pair of mixed states $\rho$ and $\sigma$ is accepted
in the controlled-swap test is exactly
${\frac{1}{2} + \frac{1}{2} \tr (\rho \sigma)}$.
\label{Proposition: probability accepted by controlled-swap test}
\end{proposition}

\begin{proof}
Let $\calR_1$ and $\calR_2$ denote the Hilbert spaces
corresponding to $\sfR_1$ and $\sfR_2$, respectively.

Let ${\rho = \sum_i p_i \ketbra{e_i}}$
and ${\sigma = \sum_i q_i \ketbra{f_i}}$
be the decompositions of $\rho$ and $\sigma$
with respect to some orthonormal bases
$\{\ket{e_i}\}$ and $\{\ket{f_i}\}$
of $\calR_1$ and $\calR_2$, respectively.
Then the state in ${(\sfR_1, \sfR_2)}$ is
${\ket{e_i} \otimes \ket{f_j}}$ with probability ${p_i q_j}$,
and in such a case,
the test results in acceptance with probability
${\frac{1}{2} + \frac{\abs{\braket{e_i}{f_j}}^2}{2}}$.

Therefore, the states $\rho$ and $\sigma$ are accepted with probability
\[
\begin{split}
\sum_i \sum_j
  p_i q_j
  \left(
    \frac{1}{2} + \frac{\abs{\braket{e_i}{f_j}}^2}{2}
  \right)
&
=
\frac{1}{2}
+
\frac{1}{2}
\sum_i \sum_j
  p_i q_j
  \braket{e_i}{f_j}\braket{f_j}{e_i}
\\
&
=
\frac{1}{2}
+
\frac{1}{2}
\sum_i \sum_j
  p_i q_j
  \tr(\ket{e_i}\braket{e_i}{f_j}\bra{f_j})
\\
&
=
\frac{1}{2}
+
\frac{1}{2}
\tr
\biggl[
  \Bigl(
    \sum_i p_i \ketbra{e_i}
  \Bigr)
  \Bigl(
    \sum_j q_j \ketbra{f_j}
  \Bigr)
\biggr]
\\
&
=
\frac{1}{2}
+
\frac{1}{2} \tr (\rho \sigma),
\end{split}
\]
as desired.
\end{proof}

% ---------------------------------------------------------------------------
%   A Key Lemma
% ---------------------------------------------------------------------------

\subsection{Reducing the Number of Proofs}

Using Proposition~\ref{Proposition: probability accepted by controlled-swap test},
we can show the following lemma,
which is the key to proving Theorem~\ref{Theorem: necessary and sufficient condition for QMA(k, c, s) = QMA(2, 2/3, 1/3)}.

\begin{lemma}
For any polynomially bounded function
$\function{k}{\Nonnegative}{\Natural}$,
any ${r \in \{0, 1, 2\}}$,
and any functions $\function{\varepsilon, \delta}{\Nonnegative}{[0,1]}$
satisfying ${\delta > 10 \varepsilon}$,
\[
  \QMA(3k+r, 1 - \varepsilon, 1 - \delta)
  \subseteq
  \QMA \Bigl(
         2k+r, 1 - \frac{\varepsilon}{2}, 1 - \frac{\delta}{20}
       \Bigr).
\]
\label{Lemma: 3k to 2k}
\end{lemma}

The essence of the proof of Lemma~\ref{Lemma: 3k to 2k}
is the basis case where ${k=1}$ and ${r=0}$.
We first give a proof for this particular case,
which will be helpful to see the idea.

\begin{proposition}
For any functions $\function{\varepsilon, \delta}{\Nonnegative}{[0,1]}$
satisfying ${\delta > 10 \varepsilon}$,
\[
  \QMA(3, 1 - \varepsilon, 1 - \delta)
  \subseteq
  \QMA \Bigl(
         2, 1 - \frac{\varepsilon}{2}, 1 - \frac{\delta}{20}
       \Bigr).
\]
\label{Proposition: 3 to 2}
\end{proposition}

\begin{proof}
Let ${A = \{ \Ayes, \Ano \}}$ be a problem in
${\QMA(3, 1 - \varepsilon, 1 - \delta)}$.
Given a ${\QMA(3, 1 - \varepsilon, 1 - \delta)}$ proof system for $A$,
we construct a
${\QMA \bigl(2, 1 - \frac{\varepsilon}{2}, 1 - \frac{\delta}{20} \bigr)}$
proof system for $A$
in the following way.

Let $V$ be the quantum verifier of the original ${\QMA(3, 1 - \varepsilon, 1 - \delta)}$ proof system.
For every input $x$,
suppose that $V$ uses ${q_{\calV}(\abs{x})}$ private qubits,
and each of the quantum proofs $V$ receives consists of ${q_{\calM}(\abs{x})}$ qubits,
for some polynomially bounded functions
$\function{q_{\calV}, q_{\calM}}{\Nonnegative}{\Natural}$.
Let ${V(x)}$ be the unitary transformation $V$ applies.

Our new quantum verifier $W$ in the
${\QMA \bigl(2, 1 - \frac{\varepsilon}{2}, 1 - \frac{\delta}{20} \bigr)}$
proof system prepares quantum registers
$\sfR_1$, $\sfR_2$, $\sfS_1$, and $\sfS_2$
for quantum proofs
and quantum registers $\sfV$ and $\sfB$ for his private computation.
Each $\sfR_i$ and $\sfS_i$ consists of ${q_{\calM}(\abs{x})}$ qubits,
$\sfV$ consists of ${q_{\calV}(\abs{x})}$ qubits,
and $\sfB$ consists of a single qubit.
All the qubits in ${(\sfV, \sfB)}$
are initialized to state $\ket{0}$.
$W$ receives two quantum proofs $\ket{\psi_1}$ and $\ket{\psi_2}$
of ${2q_{\calM}(\abs{x})}$ qubits
in ${(\sfR_1, \sfS_1)}$ and ${(\sfR_2, \sfS_2)}$, respectively,
which are expected to be of the form
\[
\ket{\psi_1} = \ket{\phi_1} \otimes \ket{\phi_3},\quad
\ket{\psi_2} = \ket{\phi_2} \otimes \ket{\phi_3},
\]
where each $\ket{\phi_i}$ is the $i$th quantum proof
the original quantum verifier $V$ would receive.
Of course, each $\ket{\psi_i}$ may not be of the form above
and the first and the second ${q_{\calM}(\abs{x})}$ qubits of 
$\ket{\psi_i}$ may be entangled.
Let $\calV$, $\calB$, each $\calR_i$, and each $\calS_i$
be the Hilbert spaces corresponding to the quantum registers
$\sfV$, $\sfB$, $\sfR_i$, and $\sfS_i$, respectively.

The protocol of $W$ is described in
Figure~\ref{Figure: verifier's protocol in two-proof system}.

\begin{figure}
\begin{algorithm*}{Verifier's Protocol in Two-Proof System}
\begin{step}
\item
Receive the first quantum proof $\ket{\psi_1}$ in ${(\sfR_1, \sfS_1)}$ 
and the second quantum proof $\ket{\psi_2}$ in ${(\sfR_2, \sfS_2)}$.
\item
Do one of the following two tests uniformly at random.
\begin{step}
\item
(\textsc{Separability test})\\
Perform the controlled-swap test over $\sfS_1$ and $\sfS_2$
using $\sfB$ as a control qubit.
That is, perform the following:
\begin{step}
\item
  Apply the Hadamard transformation $H$ to $\sfB$.
\item
  Apply the controlled-swap operator to $\sfS_1$ and $\sfS_2$
  using $\sfB$ as a control qubit.
\item
  Apply the Hadamard transformation $H$ to $\sfB$.
  Accept if $\sfB$ contains $0$, and reject otherwise.
\end{step}
\item
(\textsc{Consistency test})\\
Apply ${V(x)}$ to the qubits in ${(\sfV, \sfR_1, \sfR_2, \sfS_1)}$.
Accept iff the result corresponds to
the accepting computation of the original quantum verifier.
\end{step}
\end{step}
\end{algorithm*}
\caption{Verifier's protocol in two-proof system.}
\label{Figure: verifier's protocol in two-proof system}
\end{figure}

For the completeness, suppose that the input $x$ is in $\Ayes$.
In the original proof system,
there exist quantum proofs $\ket{\phi_1}$, $\ket{\phi_2}$, and $\ket{\phi_3}$
that cause the original quantum verifier $V$ to accept $x$
with probability at least ${1 - \varepsilon(\abs{x})}$.
In the constructed protocol,
let the quantum proofs $\ket{\psi_1}$ and $\ket{\psi_2}$ be of the form
${\ket{\psi_1} = \ket{\phi_1} \otimes \ket{\phi_3}}$
and ${\ket{\psi_2} = \ket{\phi_2} \otimes \ket{\phi_3}}$.
Then it is obvious that the constructed quantum verifier $W$
accepts $x$ with certainty in the \textsc{Separability test}
and with probability at least ${1 - \varepsilon(\abs{x})}$
in the \textsc{Consistency test},
and thus, the completeness follows.

Now for the soundness,
assume that the input $x$ is in $\Ano$.

Consider any pair of quantum proofs $\ket{\psi'_1}$ and $\ket{\psi'_2}$
of ${2q_{\calM}(\abs{x})}$ qubits,
which are set in the pairs of the quantum registers
${(\sfR_1, \sfS_1)}$ and ${(\sfR_2, \sfS_2)}$, respectively.
Let ${\rho = \tr_{\calR_1} \ketbra{\psi'_1}}$
and ${\sigma = \tr_{\calR_2} \ketbra{\psi'_2}}$.

\begin{itemize}
\item[(i)]
In the case ${\tr (\rho \sigma) \leq 1 - \frac{\delta}{5}}$:\\
In this case,
by Proposition~\ref{Proposition: probability accepted by controlled-swap test},
the probability $p_{\text{sep}}$
that the input $x$ is accepted in the \textsc{Separability test}
is at most
\[
p_{\text{sep}}
\leq
\frac{1}{2} + \frac{1}{2} \Bigl(1 - \frac{\delta}{5} \Bigr)
=
1 - \frac{\delta}{10}.
\]
Thus the verifier $W$ accepts the input $x$
with probability at most
${
\frac{1}{2} + \frac{p_{\text{sep}}}{2}
\leq 1 - \frac{\delta}{20}
}$.
\item[(ii)]
In the case ${\tr (\rho \sigma) > 1 - \frac{\delta}{5}}$:\\
Let ${\widetilde{V} = V(x) \otimes I_{\calS_2}}$
and ${\widetilde{\Pi}_{\acc} = \Pi_{\acc} \otimes I_{\calS_2}}$,
where $\Pi_{\acc}$ is the projection onto accepting states
of the original proof system.
For notational convenience,
here it is assumed that
$\widetilde{V}$ and $\widetilde{\Pi}_{\acc}$ are applied to
${(\sfV, \sfR_1, \sfS_1, \sfR_2, \sfS_2)}$
in this order of registers,
although the registers to which ${V(x)}$ and $\Pi_{\acc}$ are applied
are assumed to be in order of $\sfV$, $\sfR_1$, $\sfR_2$, and $\sfS_1$.
%% although we have defined ${V(x)}$ as being applied to
%% ${(\sfV, \sfR_1, \sfR_2, \sfS_1)}$
%% in this order of registers.
Let
${
  \ket{\alpha}
  =
  \frac{1}
       {
	 \normL{
	         \widetilde{\Pi}_{\acc}
		 \widetilde{V}
		 \left(
		   \ket{0_{\calV}} \otimes \ket{\psi'_1} \otimes \ket{\psi'_2}
		 \right)
               }
       }
  \widetilde{\Pi}_{\acc}
  \widetilde{V}
  \bigl( \ket{0_{\calV}} \otimes \ket{\psi'_1} \otimes \ket{\psi'_2} \bigr)
}$
and
${\ket{\beta} = \ket{0_{\calV}} \otimes \ket{\psi'_1} \otimes \ket{\psi'_2}}$.
Then the probability $p_{\text{cons}}$
that the input $x$ is accepted in the \textsc{Consistency test}
is at most
\[
p_{\text{cons}} \leq
F \bigl(
    \conjugate{\widetilde{V}} \ketbra{\alpha} \widetilde{V}, \ketbra{\beta}
  \bigr)^2.
\]

The fact ${\tr (\rho \sigma) > 1 - \frac{\delta}{5}}$
implies that the maximum eigenvalue $\lambda$ of $\rho$ satisfies
${\lambda > 1 - \frac{\delta}{5}}$.
Thus there exists a pure state
${\ket{\phi'_1} \in \calR_1 \otimes \calS_1}$
of the form
${\ket{\phi'_1} = \ket{\xi'_1} \otimes \ket{\eta'_1}}$
for some pure states ${\ket{\xi'_1} \in \calR_1}$
and ${\ket{\eta'_1} \in \calS_1}$
such that
\[
F \bigl( \ketbra{\phi'_1}, \ketbra{\psi'_1} \bigr)
>
\sqrt{1 - \frac{\delta}{5}},
\]
since ${\rho = \tr_{\calR_1} \ketbra{\psi'_1}}$.
Similarly, the maximum eigenvalue of $\sigma$
is more than ${1 - \frac{\delta}{5}}$
and there exists a pure state
${\ket{\phi'_2} \in \calR_2 \otimes \calS_2}$
of the form
${\ket{\phi'_2} = \ket{\xi'_2} \otimes \ket{\eta'_2}}$
for some pure states
${\ket{\xi'_2} \in \calR_2}$
and ${\ket{\eta'_2} \in \calS_2}$
such that
\[
F \bigl( \ketbra{\phi'_2}, \ketbra{\psi'_2} \bigr)
>
\sqrt{1 - \frac{\delta}{5}}.
\]
Therefore, letting
${
  \ket{\gamma}
  =
   \ket{0_{\calV}} \otimes \ket{\phi'_1} \otimes \ket{\phi'_2}
%%    \ket{0_{\calV}} \otimes \ket{\phi'_1} \otimes \ket{\phi'_3} \otimes \ket{\phi'_2} \otimes \ket{\phi'_4}
}$,
we have from Lemma~\ref{Lemma: product of fidelity} that
\[
F(\ketbra{\beta}, \ketbra{\gamma}) > 1 - \frac{\delta}{5}.
\]
Furthermore, from the soundness condition of the original proof system,
it is easy to see that
\[
F \bigl(
    \conjugate{\widetilde{V}} \ketbra{\alpha} \widetilde{V}, \ketbra{\gamma}
  \bigr)
=
F \bigl(
    \ketbra{\alpha}, \widetilde{V} \ketbra{\gamma} \conjugate{\widetilde{V}}
  \bigr)
\leq
\sqrt{1 - \delta}.
\]

Using Lemma~\ref{Lemma: F(a,b)^2 + F(b,c)^2 < 1 + F(a,c)},
we have that
\[
F \bigl(
    \conjugate{\widetilde{V}} \ketbra{\alpha} \widetilde{V}, \ketbra{\beta}
  \bigr)^2
+
F(\ketbra{\beta}, \ketbra{\gamma})^2
\leq
1
+
F \bigl(
    \conjugate{\widetilde{V}} \ketbra{\alpha} \widetilde{V}, \ketbra{\gamma}
  \bigr).
\]
It follows that
\[
\begin{split}
p_{\text{cons}}
&
\leq
1
+
F \bigl(
    \conjugate{\widetilde{V}} \ketbra{\alpha} \widetilde{V}, \ketbra{\gamma}
  \bigr)
-
F(\ketbra{\beta}, \ketbra{\gamma})^2
\\
&
<
1 + \sqrt{1 - \delta} - \Bigl( 1 - \frac{\delta}{5} \Bigr)^2
\leq
2 - \frac{\delta}{2} - 1 + \frac{2\delta}{5} - \frac{\delta^2}{25}
\leq
1 - \frac{\delta}{10}.
\end{split}
\]

Thus the verifier $W$ accepts the input $x$
with probability at most
${\frac{1}{2} + \frac{p_{\text{cons}}}{2} \leq 1 - \frac{\delta}{20}}$.
\end{itemize}
Hence the soundness is at most ${1 - \frac{\delta}{20}}$, as required.
\end{proof}

Now we prove Lemma~\ref{Lemma: 3k to 2k}.

\begin{proof}[Proof of Lemma~\ref{Lemma: 3k to 2k}]
The proof is a simple generalization of the case of Proposition~\ref{Proposition: 3 to 2}.

Let ${A = \{ \Ayes, \Ano \}}$ be a problem in
${\QMA(3k+r, 1 - \varepsilon, 1 - \delta)}$.
Given a ${\QMA(3k+r, 1 - \varepsilon, 1 - \delta)}$ proof system for $A$,
we construct a
${\QMA \bigl(2k+r, 1 - \frac{\varepsilon}{2}, 1 - \frac{\delta}{20} \bigr)}$
proof system for $A$
in the following way.

Let $V$ be the quantum verifier of the original ${\QMA(3k+r, 1 - \varepsilon, 1 - \delta)}$ proof system.
For every input $x$,
suppose that $V$ uses ${q_{\calV}(\abs{x})}$ private qubits,
and each of quantum proofs $V$ receives consists of ${q_{\calM}(\abs{x})}$ qubits,
for some polynomially bounded functions
$\function{q_{\calV}, q_{\calM}}{\Nonnegative}{\Natural}$.
Let ${V(x)}$ be the unitary transformation $V$ applies.

Our new quantum verifier $W$ in the
${\QMA \bigl(2k+r, 1 - \frac{\varepsilon}{2}, 1 - \frac{\delta}{20} \bigr)}$
proof system prepares quantum registers
${\sfR_{1,1}, \ldots, \sfR_{1,k}}$,
${\sfR_{2,1}, \ldots, \sfR_{2,k}}$,
${\sfS_{1,1}, \ldots, \sfS_{1,k}}$,
${\sfS_{2,1}, \ldots, \sfS_{2,k}}$,
${\sfR_{3,1}, \ldots, \sfR_{3,r}}$,
${\sfS_{3,1}, \ldots, \sfS_{3,r}}$
for quantum proofs
and quantum registers $\sfV$ and $\sfB$ for his private computation.
Each of $\sfR_{i,j}$ and $\sfS_{i,j}$ consists of ${q_{\calM}(\abs{x})}$ qubits,
$\sfV$ consists of ${q_{\calV}(\abs{x})}$ qubits,
and $\sfB$ consists of a single qubit.
All the qubits in ${(\sfV, \sfB)}$
are initialized to state $\ket{0}$.
Let $\calV$, $\calB$, each $\calR_{i,j}$, and each $\calS_{i,j}$
be the Hilbert spaces corresponding to the quantum registers
$\sfV$, $\sfB$, $\sfR_{i,j}$, and $\sfS_{i,j}$, respectively.
$W$ receives ${2k+r}$ quantum proofs
${\ket{\psi_{1,1}}, \ldots, \ket{\psi_{1,k}}}$,
${\ket{\psi_{2,1}}, \ldots, \ket{\psi_{2,k}}}$,
and
${\ket{\psi_{3,1}}, \ldots, \ket{\psi_{3,r}}}$
of ${2q_{\calM}(\abs{x})}$ qubits
in ${(\sfR_{1,1}, \sfS_{1,1}), \ldots, (\sfR_{1,k}, \sfS_{1,k})}$,
${(\sfR_{2,1}, \sfS_{2,1}), \ldots, (\sfR_{2,k}, \sfS_{2,k})}$,
and ${(\sfR_{3,1}, \sfS_{3,1}), \ldots, (\sfR_{3,r}, \sfS_{3,r})}$,
respectively,
which are expected to be of the form
\begin{align*}
\ket{\psi_{1,j_1}} & = \ket{\phi_{j_1}} \otimes \ket{\phi_{2k+j_1}},\\
\ket{\psi_{2,j_1}} & = \ket{\phi_{k+j_1}} \otimes \ket{\phi_{2k+j_1}},\\
\ket{\psi_{3,j_2}} & = \ket{\phi_{3k+j_2}} \otimes \ket{0_{\calS_{3,j_2}}},
\end{align*}
for each ${1 \leq j_1 \leq k}$ and ${1 \leq j_2 \leq r}$,
where each $\ket{\phi_i}$ is the $i$th quantum proof
the original quantum verifier $V$ would receive.
%% and $\calS_{i,j}$ is the Hilbert space
%% corresponding to the quantum register $\sfS_{i,j}$ for each $i$ and $j$.

The protocol of $W$ is described in
Figure~\ref{Figure: verifier's protocol in (2k+r)-proof system}.

\begin{figure}
\begin{algorithm*}{\boldmath{Verifier's Protocol in ${(2k+r)}$-Proof System}}
\begin{step}
\item
For each
${
  (i,j)
  \in
  \{(1,1), \ldots, (1,k), (2,1), \ldots, (2,k), (3,1), \ldots, (3,r)\}
}$,
receive the quantum proof $\ket{\psi_{i,j}}$ in ${(\sfR_{i,j}, \sfS_{i,j})}$.
Reject if any of the qubits in $\sfS_{3,j}$ contains $1$,
for ${1 \leq j \leq r}$.
\item
Do one of the following two tests uniformly at random.
\begin{step}
\item
(\textsc{Separability test})\\
Perform the controlled-swap test over
${(\sfS_{1,1}, \ldots, \sfS_{1,k})}$
and ${(\sfS_{2,1}, \ldots, \sfS_{2,k})}$
using $\sfB$ as a control qubit.
That is, perform the following:
\begin{step}
\item
  Apply the Hadamard transformation $H$ to $\sfB$.
\item
  Apply the controlled-swap operator to
  ${(\sfS_{1,1}, \ldots, \sfS_{1,k})}$
  and ${(\sfS_{2,1}, \ldots, \sfS_{2,k})}$
  using $\sfB$ as a control qubit.
\item
  Apply the Hadamard transformation $H$ to $\sfB$.
  Accept if $\sfB$ contains $0$, and reject otherwise.
\end{step}
\item
(\textsc{Consistency test})\\
Apply ${V(x)}$ to the qubits in
${
  (
    \sfV,
    \sfR_{1,1}, \ldots, \sfR_{1,k},
    \sfR_{2,1}, \ldots, \sfR_{2,k},
    \sfS_{1,1}, \ldots, \sfS_{1,k},
    \sfR_{3,1}, \ldots, \sfR_{3,r}
  )
}$.
Accept iff the result corresponds to
the accepting computation of the original quantum verifier.
\end{step}
\end{step}
\end{algorithm*}
\caption{Verifier's protocol in ${(2k+r)}$-proof system.}
\label{Figure: verifier's protocol in (2k+r)-proof system}
\end{figure}

The rest of the proof is essentially the same as
in the case of Proposition~\ref{Proposition: 3 to 2}.
%% Now the claim follows from
%% the argument almost parallel to the proof of Proposition~\ref{Proposition: 3 to 2}.
When analyzing soundness,
consider any set of ${2k+r}$ quantum proofs
${\ket{\psi'_{1,1}}, \ldots, \ket{\psi'_{1,k}}}$,
${\ket{\psi'_{2,1}}, \ldots, \ket{\psi'_{2,k}}}$,
and
${\ket{\psi'_{3,1}}, \ldots, \ket{\psi'_{3,r}}}$
of ${2q_{\calM}(\abs{x})}$ qubits,
which are set in the quantum registers
${(\sfR_{1,1}, \sfS_{1,1}), \ldots, (\sfR_{1,k}, \sfS_{1,k})}$,
${(\sfR_{2,1}, \sfS_{2,1}), \ldots, (\sfR_{2,k}, \sfS_{2,k})}$,
and ${(\sfR_{3,1}, \sfS_{3,1}), \ldots, (\sfR_{3,r}, \sfS_{3,r})}$,
respectively,
and let
${
  \ket{\psi'_1}
  =
  \ket{\psi'_{1,1}} \otimes \cdots \otimes \ket{\psi'_{1,k}}
}$
and
${
  \ket{\psi'_2}
  =
  \ket{\psi'_{2,1}} \otimes \cdots \otimes \ket{\psi'_{2,k}}
}$.
Let
${\rho = \rho_1 \otimes \cdots \otimes \rho_k}$
and ${\sigma = \sigma_1 \otimes \cdots \otimes \sigma_k}$,
where ${\rho_j = \tr_{\calR_{1,j}} \ketbra{\psi'_{1,j}}}$
and ${\sigma_j = \tr_{\calR_{2,j}} \ketbra{\psi'_{2,j}}}$
for each ${1 \leq j \leq k}$.
Note that,
if ${\tr (\rho \sigma) > 1 - \frac{\delta}{5}}$,
there exist pure states
${\ket{\xi'_{1,j}} \in \calR_{1,j}}$,
${\ket{\xi'_{2,j}} \in \calR_{2,j}}$,
${\ket{\eta'_{1,j}} \in \calS_{1,j}}$,
and ${\ket{\eta'_{2,j}} \in \calS_{2,j}}$
for each ${1 \leq j \leq k}$
such that
the states
${
  \ket{\phi'_1}
  =
  \ket{\xi'_{1,1}} \otimes \ket{\eta'_{1,1}}
  \otimes \cdots \otimes
  \ket{\xi'_{1,k}} \otimes \ket{\eta'_{1,k}}
}$
and
${
  \ket{\phi'_2}
  =
  \ket{\xi'_{2,1}} \otimes \ket{\eta'_{2,1}}
  \otimes \cdots \otimes
  \ket{\xi'_{2,k}} \otimes \ket{\eta'_{2,k}}
}$
satisfy that
${
  F \bigl( \ketbra{\phi'_1}, \ketbra{\psi'_1} \bigr)
  >
  \sqrt{1 - \frac{\delta}{5}}
}$
and
${
  F \bigl( \ketbra{\phi'_2}, \ketbra{\psi'_2} \bigr)
  >
  \sqrt{1 - \frac{\delta}{5}}
}$.
Now the claim follows from
the argument almost parallel to the proof of Proposition~\ref{Proposition: 3 to 2}.
\end{proof}

Now Theorem~\ref{Theorem: necessary and sufficient condition for QMA(k, c, s) = QMA(2, 2/3, 1/3)}
can be proved
by using the transformation in Lemma~\ref{Lemma: 3k to 2k} repeatedly.

\begin{proof}[Proof of Theorem~\ref{Theorem: necessary and sufficient condition for QMA(k, c, s) = QMA(2, 2/3, 1/3)}]
The ``only if'' part is obvious and we show the ``if'' part.

From Corollary~\ref{Corollary: exp-small completeness error for QMA},
we have that
${
  \QMA(k, c, s)
  \subseteq
  \QMA \bigl(k, 1 - 2^{-p}, 1 - \frac{c-s}{2} \bigr)
}$
for any polynomially bounded function
$\function{p}{\Nonnegative}{\Natural}$.
Now we repeatedly apply the transformation in Lemma~\ref{Lemma: 3k to 2k}
${O(\log k)}$ times,
and finally we can show the inclusion
${
  \QMA \bigl(k, 1 - 2^{-p}, 1 - \frac{c-s}{2} \bigr)
  \subseteq
  \QMA \bigl(2, 1 - 2^{-p}, 1 - \frac{1}{q} \bigr)
}$
for some polynomially bounded function
$\function{q}{\Nonnegative}{\Natural}$.
Note that the size of the circuit of the verifier
after each application of the transformation in Lemma~\ref{Lemma: 3k to 2k}
is at most some constant times that of the original verifier
plus an amount bounded by a polynomial in the input length.
Thus, given a description of the circuit of the verifier
in the original $k$-proof system,
one can compute in time polynomial in the input length
a description of the circuit of the verifier
in the resulting two-proof system.
From our assumption, 
${
  \QMA \bigl(2, 1 - 2^{-p}, 1 - \frac{1}{q} \bigr)
  =
  \QMA \bigl(2, \frac{2}{3}, \frac{1}{3} \bigr)
}$,
and thus, the inclusion
${
  \QMA(k, c, s)
  \subseteq
  \QMA \bigl(2, \frac{2}{3}, \frac{1}{3} \bigr)
}$
follows.
The other inclusion is trivial
since our assumption implies that
 ${
  \QMA \bigl(2, \frac{2}{3}, \frac{1}{3} \bigr)
  =
  \QMA(2, c, s)
}$,
and we have the theorem.
\end{proof}

The following is an immediate corollary of
Theorem~\ref{Theorem: necessary and sufficient condition for QMA(k, c, s) = QMA(2, 2/3, 1/3)}.

\begin{corollary}
${\QMA(k, c, s) = \QMA}$
for any polynomially-bounded function
$\function{k}{\Nonnegative}{\Natural}$ satisfying ${k \geq 2}$
and any two-sided bounded error probability ${(c, s)}$
if and only if
${\QMA(2, c, s) = \QMA}$
for any two-sided bounded error probability ${(c, s)}$.
\label{Corollary: condition under which QMA(k, c, s) = QMA}
\end{corollary}

\begin{proof}
The proof is almost parallel to the proof of
Theorem~\ref{Theorem: necessary and sufficient condition for QMA(k, c, s) = QMA(2, 2/3, 1/3)}.
Again the ``only if'' part is obvious and we show the ``if'' part.

Using the same argument as in the proof of
Theorem~\ref{Theorem: necessary and sufficient condition for QMA(k, c, s) = QMA(2, 2/3, 1/3)},
we can show the inclusion
${
  \QMA(k, c, s)
  \subseteq
  \QMA \bigl(2, 1 - 2^{-p}, 1 - \frac{1}{q} \bigr)
}$
for some polynomially bounded function
$\function{q}{\Nonnegative}{\Natural}$.
From our assumption, 
${
  \QMA \bigl(2, 1 - 2^{-p}, 1 - \frac{1}{q} \bigr)
  =
  \QMA
}$,
and thus, the inclusion ${\QMA(k, c, s) \subseteq \QMA}$ follows.
The other inclusion is trivial
since our assumption implies that ${\QMA = \QMA(2, c, s)}$,
and we have the corollary.
\end{proof}

\begin{remark}
Theorem~\ref{Theorem: necessary and sufficient condition for QMA(k, c, s) = QMA(2, 2/3, 1/3)}
improves the original statement in our conference version~\cite{KobMatYam03ISAAC}
in two ways.
First, the condition~($\ast$) now only requires
the amplifiablity of the success probability for two-proof systems,
whereas our original condition required it for every $k$-proof system.
Second, now ${\QMA(k)}$ even with every polynomially-bounded function $k$
coincides with ${\QMA(2)}$ if the condition holds.
Previously, we showed it only for ${\QMA(k)}$ with every \emph{constant} $k$.
The same improvements were independently done
by Aaronson,~Beigi,~Drucker,~Fefferman,~and~Shor~\cite{AarBeiDruFefSho08CCC}
but with a different proof.
Instead of repeatedly applying the transformation
that reduces the number of proofs by two-thirds as above,
they showed a direct method of reducing the number of proofs to two~\cite[Theorem~23]{AarBeiDruFefSho08CCC}.
Although the resulting two-proof system from their transformation
also has soundness only polynomially bounded away from one,
their soundness is better than ours in most cases
(except for the case where the gap between completeness $c$ and soundness $s$
 in the original system is so small relative to the number $k$ of proofs that
 ${c-s \in o(k^{-\alpha})}$,
 where ${\alpha = \frac{\log 20}{\log 3 - 1} - 1 \approx 6.388\cdots}$,
 in which case our analysis gives better soundness).
%%  ${c-s \in o(k^{-7})}$, in which case our analysis gives better soundness).
\end{remark}

% ---------------------------------------------------------------------------
%   Cases with Perfect Soundness
% ---------------------------------------------------------------------------

\section{Cases with Perfect Soundness}
\label{Section: cases with perfect soundness}

This section focuses on the quantum multi-proof Merlin-Arthur proof systems
of perfect soundness.
In the case of perfect soundness,
it is proved that
multiple quantum proofs do not increase the verification power,
which also gives a connection
between two existing concepts of ``quantum nondeterminism''.
Formally, the following is proved.

\begin{theorem}
For any polynomially bounded function
$\function{k}{\Nonnegative}{\Natural}$
and any function $\function{c}{\Nonnegative}{[0,1]}$,
\[
\QMA(k, c, 0) = \QMA(1, c, 0).
\]
\label{Theorem: QMA(k,c,0)=QMA(1,c,0)}
\end{theorem}

\begin{proof}
Let ${A = \{ \Ayes, \Ano \}}$ be a problem in ${\QMA(k, c, 0)}$.
Given a ${\QMA(k, c, 0)}$ proof system for $A$,
we construct a ${\QMA(1, c, 0)}$ proof system for $A$ in the following way.

Let $V$ be a quantum verifier of the ${\QMA(k, c, 0)}$ proof system.
For every input $x$,
assume that each quantum proof $V$ receives is of size ${q(\abs{x})}$.

Our new quantum verifier $W$ in the ${\QMA(1, c, 0)}$ proof system
receives one quantum proof of size ${k(\abs{x})q(\abs{x})}$
and simulates $V$ with this quantum proof.

The completeness is clearly at least $c$.

For the soundness, assume that the input $x$ is in $\Ano$.
Let $\ket{\phi}$ be any quantum proof of size ${k(\abs{x})q(\abs{x})}$.
Let $e_i$ be the lexicographically $i$th string in $\Sigma^{k(\abs{x})q(\abs{x})}$.
Note that, for every $i$,
the original verifier $V$ never accepts $x$
when the ${k(\abs{x})}$ quantum proofs he receives form the state $\ket{e_i}$.
Since any $\ket{\phi}$ is expressed as a linear combination of these $\ket{e_i}$,
it follows that $W$ rejects $x$ with certainty.
\end{proof}

Let ${\EQMA(k) = \QMA(k,1,0)}$
and ${\RQMA(k) = \QMA \bigl(k, \frac{1}{2}, 0 \bigr)}$ for every $k$.
Theorem~\ref{Theorem: QMA(k,c,0)=QMA(1,c,0)}
implies that ${\EQMA(k) = \EQMA(1)}$ and ${\RQMA(k) = \RQMA(1)}$.
Furthermore, one can consider
the complexity class ${\NQMA(k)}$
that combines two existing concepts of ``quantum nondeterminism'',
${\QMA(k)}$ and $\NQP$.

\begin{definition}
A problem ${A = \{ \Ayes, \Ano \}}$ is in ${\NQMA(k)}$
if there exists a function
$\function{c}{\Nonnegative}{(0,1]}$
such that $A$ is in ${\QMA(k, c, 0)}$.
\label{Definition: NQMA}
\end{definition}
Note that ${\NQMA(k) = \NQMA(1)}$ is also immediate
from Theorem~\ref{Theorem: QMA(k,c,0)=QMA(1,c,0)}.
The next theorem shows that ${\NQMA(1)}$ coincides with
the class $\NQP$.

\begin{theorem}
${\EQMA(1) \subseteq \RQMA(1) \subseteq \NQMA(1) = \NQP}$.
\label{Theorem: NQMA = NQP}
\end{theorem}

\begin{proof}
It is sufficient to show that ${\NQMA(1) \subseteq \NQP}$,
since ${\EQMA(1) \subseteq \RQMA(1) \subseteq \NQMA(1)}$
and ${\NQMA(1) \supseteq \NQP}$ hold obviously.

Let ${A = \{ \Ayes, \Ano \}}$ be a problem in ${\NQMA(1)}$.
Given an ${\NQMA(1)}$ proof system for $A$,
we construct an $\NQP$ algorithm for $A$.

Let $V$ be the quantum verifier of the ${\NQMA(1)}$ proof system.
For every input $x$,
suppose that $V$ uses ${q_{\calV}(\abs{x})}$ private qubits,
and each quantum proof $V$ receives consists of ${q_{\calM}(\abs{x})}$ qubits,
for some polynomially bounded functions
$\function{q_{\calV}, q_{\calM}}{\Nonnegative}{\Natural}$.
Let ${V(x)}$ be the unitary transformation $V$ applies.

In the $\NQP$ algorithm for $A$,
we prepare quantum registers $\sfR$, $\sfS_1$, and $\sfS_2$,
where $\sfR$ consists of ${q_{\calV}(\abs{x})}$ qubits
and each $\sfS_i$ consists of ${q_{\calM}(\abs{x})}$ qubits.
All the qubits in $\sfR$, $\sfS_1$, and $\sfS_2$
are initialized to state $\ket{0}$.
The precise algorithm is described in
Figure~\ref{Figure: NQP simulation of NQMA proof system}.

\begin{figure}
\begin{algorithm*}{\boldmath{$\NQP$ Simulation of $\NQMA$ Proof System}}
\begin{step}
\item
Apply the Hadamard transformation $H$ to every qubit in $\sfS_1$.
\item
Copy the contents of $\sfS_1$ to those of $\sfS_2$.
\item
Apply ${V(x)}$ to the pair of quantum registers ${(\sfR, \sfS_1)}$.
Accept if the contents of ${(\sfR, \sfS_1)}$
make the original verifier accept.
\end{step}
\end{algorithm*}
\caption{$\NQP$ simulation of an $\NQMA$ proof system.}
\label{Figure: NQP simulation of NQMA proof system}
\end{figure}

For the completeness, suppose that the input $x$ is in $\Ayes$.
In the original ${\NQMA(1)}$ proof system for $A$,
there exists a quantum proof $\ket{\phi}$ of size ${q_{\calM}(\abs{x})}$
that causes $V$ to accept $x$ with non-zero probability.
Suppose that $V$ never accepts $x$
with any given quantum proof $\ket{e_i}$
for ${1 \leq i \leq 2^{q_{\calM}(\abs{x})}}$,
where $e_i$ is the lexicographically $i$th string in $\Sigma^{q_{\calM}(\abs{x})}$.
Then with a similar argument to the proof of
Theorem~\ref{Theorem: QMA(k,c,0)=QMA(1,c,0)},
$V$ never accepts $x$
with any given quantum proof $\ket{\phi}$ of size ${q_{\calM}(\abs{x})}$,
which contradicts the assumption.
Thus there is at least one $\ket{e_i}$
that causes $V$ to accept $x$ with non-zero probability.
Hence, in the algorithm in Figure~\ref{Figure: NQP simulation of NQMA proof system},
the probability of acceptance must be non-zero,
since it simulates with probability $2^{-q_{\calM}(\abs{x})}$
the case where $V$ is given a proof $\ket{e_i}$ for every $i$.

Now for the soundness, suppose that the input $x$ is in $\Ano$.
In the original ${\NQMA(1)}$ proof system for $A$,
no matter which quantum proof $\ket{\phi}$ of size ${q_{\calM}(\abs{x})}$ is given,
$V$ never accepts $x$.
Hence, in the algorithm in Figure~\ref{Figure: NQP simulation of NQMA proof system},
the probability of acceptance is zero
and the soundness follows.
\end{proof}

Now the following characterization of $\NQP$ is immediate.
%from Theorem~\ref{Theorem: NQMA = NQP}.

\begin{corollary}
\label{Corollary: characterization of NQP}
${\NQP= \bigcup_{\function{c}{\Nonnegative}{(0,1]}} \QMA(1,c,0)}$.
\end{corollary}

% ---------------------------------------------------------------------------
%   Discussions
% ---------------------------------------------------------------------------

\section{Discussions}
\label{Section: discussions}

This section shows that there is no POVM measurement
that determines whether a given unknown state is in a tensor product form
or even maximally entangled.

Suppose that there is a quantum subroutine that
answers which of the following~(a)~and~(b) is true for a given proof
${\ket{\Psi} \in \calH^{\otimes 2}}$ of $2n$ qubits,
where $\calH$ is the Hilbert space consisting of $n$ qubits:
\begin{itemize}
\item[(a)]
$\ketbra{\Psi}$ is in
${
  \sfH_0
  = 
  \bigset{\ketbra{\Psi_0}}
	 {
	   \ket{\Psi_0} \in \calH^{\otimes 2},
	   \;
	   \exists \ket{\phi}, \ket{\psi} \in \calH,
	   \;
	   \ket{\Psi_0} = \ket{\phi} \otimes \ket{\psi}
	 }
}$,
\item[(b)]
$\ketbra{\Psi}$ is in
${
  \sfH^{\varepsilon}_1
  =
  \bigset{\ketbra{\Psi_1}}
	 {
	   \ket{\Psi_1} \in \calH^{\otimes 2},
	   \;
	   \max_{\ket{\phi},\ket{\psi} \in \calH}
	     F(\ketbra{\Psi_1}, \ketbra{\phi} \otimes \ketbra{\psi})
	   \leq 
	   1 - \varepsilon
	 }
}$.
\end{itemize}
As for the proof $\ket{\Psi}$ that does not satisfy (a) nor (b),
this subroutine may answer (a) or (b) arbitrarily.
The rest of this section proves that
this kind of subroutines cannot be realized by any physical method.
In fact, we prove a stronger statement
that the set of states in a tensor product form
cannot be distinguished even from the set of maximally entangled states
by any physical operation.
Here, following Ref.~\cite{BenBerPopSch96PRA},
we say that the $n$-qubit state ${\rho = \ketbra{\Psi}}$ is \emph{maximally entangled} 
if $\ket{\Psi}$ can be written as 
\[
 \ket{\Psi}
 =
 \sum_{i=1}^{d} {\alpha_i} \ket{e_i} \otimes \ket{f_i},
 \;
 \abs{\alpha_i}^2=\frac{1}{d},%\; i=1,\ldots, d,
\]
where 
${d = 2^n}$ is the dimension of $\calH$
and ${\{\ket{e_i} \}}$ and ${\{\ket{f_i} \}}$
%% and ${\{\ket{e_1}, \ldots, \ket{e_d}\}}$
%% and ${\{\ket{f_1}, \ldots, \ket{f_d}\}}$
are orthonormal bases of $\calH$.
Among all states,
maximally entangled states are farthest away from states
in a tensor product form, and
\[
\min_{\ket{\Psi} \in \calH^{\otimes 2}}
\max_{\ket{\phi}, \ket{\psi} \in \calH}
  F(\ketbra{\Psi}, \ketbra{\phi} \otimes \ketbra{\psi})
= \frac{1}{\sqrt{d}} = 2^{-\frac{n}{2}}
\]
is achieved by maximally entangled states.
%% Thus a verifier cannot rule out quantum proofs
Thus Arthur cannot rule out quantum proofs
that are far from states of a tensor product of pure states.

\begin{theorem}
Suppose that one of the following two is true for a given proof
${\ket{\Psi} \in \calH^{\otimes 2}}$ of $2n$ qubits:
\begin{itemize}
\item[(a)]
$\ketbra{\Psi}$ is in
${
  \sfH_0
  = 
  \bigset{\ketbra{\Psi_0}}
	 {
	   \ket{\Psi_0} \in \calH^{\otimes 2},
	   \;
	   \exists \ket{\phi}, \ket{\psi} \in \calH,
	   \;
	   \ket{\Psi_0} = \ket{\phi} \otimes \ket{\psi}
	 }
}$,
\item[(b)]
$\ketbra{\Psi}$ is in
${
  \sfH_1
  =
  \bigset{\ketbra{\Psi_1}}
	 {
	   \textnormal{${\ket{\Psi_1} \in \calH^{\otimes 2}}$ is maximally entangled}
	 }
}$.
\end{itemize}
Then, in determining which of~(a)~and~(b) is true,
no POVM measurement is better
than the trivial strategy in which one guesses at random without 
any operation at all.
\label{Theorem: no-distinguish}
\end{theorem}

\begin{proof}
Let ${\bmM = \{M_0, M_1\}}$ be a POVM on $\calH^{\otimes 2}$.
With $\bmM$ we conclude ${\ketbra{\Psi} \in \sfH_i}$
if $\bmM$ results in $i$, ${i \in \Binary}$.
Let $\mathrm{P}^{\bmM}_{i \rightarrow j}(\ketbra{\Psi})$
denote the probability that
${\ketbra{\Psi} \in \sfH_j}$ is concluded by $\bmM$ while 
${\ketbra{\Psi} \in \sfH_i}$ is true.
We want to find the measurement that minimizes
$\mathrm{P}^{\bmM}_{0 \rightarrow 1}(\ketbra{\Psi})$
keeping the other side of error small enough.
More precisely,
we consider $\calE$ defined and bounded as follows.
\[
\begin{split}
\calE
&
\defeq
\min_{\bmM}
\Bigset{\max_{\rho \in \sfH_0} \mathrm{P}^{\bmM}_{0 \rightarrow 1}(\rho)}
       {
	 \max_{\rho \in \sfH_1} \mathrm{P}^{\bmM}_{1 \rightarrow 0}(\rho)
	 \leq
	 \delta
       }
\\
&
\geq
\min_{\bmM}
\biggset{\int_{\rho \in \sfH_0} \mathrm{P}^{\bmM}_{0 \rightarrow 1}(\rho) \mu_0(\mathrm{d} \rho)}
	{
	  \int_{\rho \in \sfH_1} \mathrm{P}^{\bmM}_{1 \rightarrow 0}(\rho) \mu_1(\mathrm{d} \rho)
	  \leq
	  \delta
	}
\\
&
=
\min_{\bmM}
\Biggset{
          \mathrm{P}^{\bmM}_{0 \rightarrow 1}
	  \biggl(
            \int_{\rho \in \sfH_0} \rho \mu_0 (\mathrm{d} \rho)
	  \biggr)
        }
	{
	  \mathrm{P}^{\bmM}_{1 \rightarrow 0}
	  \biggl(
            \int_{\rho \in \sfH_1} \rho \mu_1 (\mathrm{d} \rho)
	  \biggr)
	  \leq
	  \delta
	},
\end{split}
\]
where each $\mu_i$ is an arbitrary probability measure in $\sfH_i$.
It follows that $\calE$ is larger than
the error probability in distinguishing
${\int_{\rho \in \sfH_0} \rho \mu_0 (\mathrm{d} \rho)}$
from ${\int_{\rho \in \sfH_1} \rho \mu_1 (\mathrm{d} \rho)}$.

Take $\mu_0$ as a uniform distribution over the set
${\{\ketbra{e_i} \otimes \ketbra{e_j}\}_{1 \leq i,j \leq d}}$,
that is, ${\mu_0 (\ketbra{e_i} \otimes \ketbra{e_j}) = \frac{1}{d^2}}$
for each $i$ and $j$,
where ${\{\ket{e_i} \}}$ is an orthonormal basis of $\calH$,
and take $\mu_1$ as a uniform distribution over the set
${\{\ketbra{g_{k,l}}\}_{1 \leq k,l \leq d}}$,
that is, ${\mu_1 (\ketbra{g_{k,l}}) = \frac{1}{d^2}}$
for each $k$ and $l$,
where 
\[
\ket{g_{k,l}}
=
\frac{1}{d}
\sum_{j=1}^d
  \bigl(
    e^{2 \pi \sqrt{-1} \frac{jk}{d}} \ket{e_j} \otimes \ket{e_{(j+l) \bmod d}}
  \bigr).
\]
This ${\{\ket{g_{k,l}} \}}$ forms an orthonormal basis of $\calH^{\otimes 2}$~\cite{BenBraCreJozPreWoo93PRL},
and thus
\[
\int_{\rho \in \sfH_0} \rho \mu_0 (\mathrm{d} \rho)
=
\int_{\rho \in \sfH_1} \rho \mu_1 (\mathrm{d} \rho)
=
\frac{1}{d^2} I_{\calH^{\otimes 2}}.
\]
Hence we have the assertion.
\end{proof}

From Theorem~\ref{Theorem: no-distinguish},
it is easy to show the following corollary.

\begin{corollary}
Suppose one of the following two is true for the proof
${\ket{\Psi} \in \calH^{\otimes 2}}$ of $2n$ qubits:
\begin{itemize}
\item[(a)]
$\ketbra{\Psi}$ is in
${
  \sfH_0
  = 
  \bigset{\ketbra{\Psi_0}}
	 {
	   \ket{\Psi_0} \in \calH^{\otimes 2},
	   \;
	   \exists \ket{\phi}, \ket{\psi} \in \calH,
	   \;
	   \ket{\Psi_0} = \ket{\phi} \otimes \ket{\psi}
	 }
}$,
\item[(b)]
$\ketbra{\Psi}$ is in
${
  \sfH^{\varepsilon}_1
  =
  \bigset{\ketbra{\Psi_1}}
	 {
	   \ket{\Psi_1} \in \calH^{\otimes 2},
	   \;
	   \max_{\ket{\phi},\ket{\psi} \in \calH}
             F(\ketbra{\Psi_1}, \ketbra{\phi} \otimes \ketbra{\psi})
	   \leq 
	   1 - \varepsilon
	 }
}$.
\end{itemize}
Then, for any ${0 \leq \varepsilon \leq 1 - 2^{-\frac{n}{2}}}$,
in determining which of~(a)~and~(b) is true,
no POVM measurement is better than
the trivial strategy in which one guesses at random without 
any operation at all.
\label{Corollary: indistinguishability}
\end{corollary}

% ---------------------------------------------------------------------------
%   Conclusions
% ---------------------------------------------------------------------------

\section{Conclusions}
\label{Section: conclusions}

This paper introduced the multi-proof version of quantum Merlin-Arthur proof systems.
To investigate the possibility that
multi-proof quantum Merlin-Arthur proof systems
collapse to usual single-proof ones,
this paper proved several basic properties
such as a necessary and sufficient condition under which
the number of quantum proofs is reducible to two.
However, the central question
whether multiple quantum proofs are indeed more helpful to Arthur
still remains open.
The authors hope that this paper sheds light on new features
on quantum Merlin-Arthur proof systems and entanglement theory,
and more widely on quantum computational complexity and
quantum information theory.

% ---------------------------------------------------------------------------
%   Acknowledgements
% ---------------------------------------------------------------------------

\subsection*{Acknowledgements}

The authors are grateful to John~Watrous for providing us
with an unpublished proof of ${\QMA \subseteq \PP}$,
which was shown jointly by Alexei~Yu.~Kitaev and John~Watrous.
Lemma~\ref{Lemma: exp-small completeness error}
was found during a discussion between HK and John~Watrous,
and HK is grateful to him.
HK thanks Richard~Cleve and Lance~Fortnow
for their helpful comments.

% ---------------------------------------------------------------------------
%   References
% ---------------------------------------------------------------------------

%% \bibliographystyle{HKplain-nonote}
%% \bibliography{QMA2}

\end{document}